\documentstyle[aps,epsfig,amssymb,multicol,citesort]{revtex}
\newcommand{\B}[1]{\mbox{\boldmath $#1$}}
 
\newcommand {\be}   {\begin{equation}}
\newcommand {\ba}   {\begin{array}}
\newcommand {\bea}  {\begin{eqnarray}}
\newcommand {\bfi}  {\begin{figure}}
\newcommand {\ee}   {\end{equation}}
\newcommand {\ea}   {\end{array}}
\newcommand {\eea}  {\end{eqnarray}}
\newcommand {\efi}  {\end{figure}}                

\newcommand {\T}   {\theta}
\newcommand {\lp}{\left(}
\newcommand {\rp}{\right)}
\newcommand {\ra}{\right\rangle}
\newcommand {\la}{\left\langle}

\begin{document}
\bibliographystyle{prsty}
\title{Intermittency in Turbulence: computing the scaling  exponents in shell  models}
\author{Roberto Benzi$^{1,2}$, Luca Biferale$^{1,2}$, 
Mauro Sbragaglia $^{1,2}$ and Federico Toschi$^{2,3}$}
\address{$^1$ Dipartimento di Fisica, Universit\`a "Tor Vergata", Via della Ricerca Scientifica 1, I-00133 Roma, Italy}
\address{$^2$ INFM, Sezione di Roma ``Tor Vergata'', Via della Ricerca Scientifica 1, I-00133 Roma, Italy}
\address{$^3$ Istituto per le Applicazioni del Calcolo, CNR, Viale del Policlinico 137, I-00161, Roma, Italy}
\maketitle
\begin{abstract}
We discuss a stochastic closure for the equation of motion satisfied
by multi-scale correlation functions in the framework of shell models
of turbulence. We give a systematic procedure to calculate the
anomalous scaling exponents of structure functions by using the exact
constraints imposed by the equation of motion. We present an explicit
calculation for fifth order scaling exponent at varying  the
free parameter entering in the non-linear term of the model.  The same
method applied to the case of shell models for Kraichnan passive
scalar provides a connection between the concept of zero-modes and
time-dependent cascade processes.
\end{abstract}
\section{Introduction}
\label{sect1} 
Since the fundamental work by Kolmogorov, it has been widely
recognized that a consistent theory for the statistical properties of
turbulence should quantitatively explain intermittency. In the last
ten years many important steps have been done in providing a
consistent picture of intermittency in turbulence. First, careful
designed experimental measurements and a systematic way to analyze data
have shown the universal feature of intermittency \cite{fri95,ben93b}.
Second, a well defined theory has been proposed to compute anomalous
scaling for a class of {\it linear} problem, i.e. the case of
Kraichnan passive scalar \cite{fgv01}.  In the latter case, the notion
of {\it zero-modes} provided a theoretical framework for many
fundamental properties of intermittency.  Yet, we are still looking
for defining a suitable strategy for a quantitative computations of
intermittency in the full non linear problem, namely the Navier-Stokes
equations. The problem of anomalous scaling must be divided in two
steps.  First, we need to clean it from all unwanted difficulties,
trying to focus on the main physical mechanism leading to small-scales
intermittency and to its connections with the non-linear structure of
the equation of motion. This is the main goal of this article. We show
that the anomalous scaling of small-scales velocity fluctuations of a
shell model of turbulence can be derived from the equation of
motion. The result is based on a stochastic closure.  A second, more
ambitious goal, is to extend this results to the full complexity of
Navier-Stokes equations. Some comments on the latter problem are also
proposed in the conclusions.

Let us make a few general comments on the nature of the problem we are
facing.  We are interested in the (universal) features of the
statistical properties of the velocity field, $\B v(\B x,t)$, in a
homogeneous and isotropic turbulent flow. Experimental data and
theoretical ideas suggest that these universal properties are related
to velocity fluctuations at scales much smaller than outer
energy input scale,  $L$. More precisely, we want to compute the simultaneous
multi-point correlation functions $C^{(n)}(\B x_1,\B x_2, \dots,\B
x_n) = \la v_{i_1}(\B x_1,t) v_{i_2}(\B x_2,t),\dots, ...v_{i_n}(\B x_n,t)
\ra$ for scale separations $|\B x_i-\B x_j|$ much smaller than
$L$. Our task must be performed by using the Navier-Stokes
equations. This is precisely the result which has been accomplished in
the case of the Kraichnan model previously referred to. The equation
of motions provide a relations among the (infinite) sets of 
{\sf simultaneous} correlation functions, $C^{(n)}(\B x_1,\B x_2,..\B x_n)$: 
\be 
0 = \frac{d}{dt} C^{(n)}(\B x_1,\B x_2,..\B x_n) = 
\Gamma[C^{(n+1)}(\B x_1;\B x_2;\dots;\B x_{n})],
\label{basic}
\ee
where we have assumed stationarity and
where with $\Gamma$ we denote the  integro-differential linear operator
 derivable explicitly  from the Navier-Stokes equations.

The equivalent of the 
above hierarchy rewritten for the case of  the Kraichnan model is closed
order by order in the correlation functions allowing for a
perturbative calculation of the statistical properties.  In the full
non linear problem of the Navier-Stokes equations, one can show that
(\ref{basic}) do not form a closed set of equations, rather it should
be considered as a constrain for the complete solution. Actually, the
fundamental quantities for studying intermittency in turbulence
involve also temporal information from {\sf multi-time} correlation functions,   $ C^{(n)}(\B x_1,t_1;\B
x_2,t_2;\dots;\B x_n,t_n) = \la v_{i_1}(\B x_1,t_1) v_{i_2}(\B
x_2,t_2)...v_{i_n}(\B x_n,t_n) \ra$. Namely, we need to look for
the solution of the problem:
\be
\frac{\partial^k}{\partial t_1 \cdots \partial t_k}
 C^{(n)}(\B x_1,t_1;\B x_2,t_2;\dots;\B x_n,t_n) =
\Sigma[C^{(n+1)}],
\label{basicns}
\ee
with $k \le n$ and where $\Sigma$ is a functional of the time
dependent correlation functions of order $n+1$ 
depending on $n+1$ velocity fields at $n$ different times. 
 The fundamental question we are
facing is which are, if any, the physical informations we should use
in order to {\it solve} (\ref{basicns}). In some broad sense, being not
able to solve (\ref{basicns}) by any kind of ``brute force'' attempt,
we still need to understand which are the correct ``order parameters''
we should consider  to find out a systematic way to compute a
solution of (\ref{basicns}).
\\  We argue that a strategy to compute the
solutions of the {\sf multi-time} hierarchy  (\ref{basicns})
 may be outlined by first finding a
``physically consistent'' solutions for the {\sf simultaneous} hierarchy
(\ref{basic}) which can be
used as the starting point for successive approximations.  For
``physically consistent'' we mean that the solution should respect the
phenomenological constraints imposed by the Navier Stokes equations and
in particular by its time-space scaling properties.  This is the main
idea pursued in this paper.  More precisely we will discuss how far we
can provide a quantitative computation of intermittency based on the
following three main points: (i) we only use the constraints coming
from the {\sf simultaneous} equations (\ref{basic}); (ii) we look for the
solutions of (\ref{basic}) by assuming that the out-of-equilibrium
statistical properties of the velocity field can be obtained by a
suitable time dependent stochastic process; (iii) we shall restrict
ourselves to non-linear shell models
\cite{fri95,bif03}.

Having discussed in details the motivation of point (i), let us
briefly comment on point (ii). Random multiplicative processes have
been often used in literature as a simple mathematical tool to
describe anomalous scaling properties of turbulent flows
\cite{ben84}. Only a few attempts have successfully linked
cascade-multiplicative process with the structure of the equation of
motion \cite{ben93}.  Recently, the concept of random multiplicative
process have been enlarged by including non trivial time dynamics
\cite{bif99,ben03}. In particular, the choice of time dynamics can be
done in order to satisfy the Navier Stokes temporal scaling (in a
Lagrangian reference frame). Moreover, it has been shown that
time dynamics affects in non trivial way also the spatial scaling of
$C^{(n)}(\B x_1,\B x_2,\dots,\B x_n)$ as a function of
intermittency. Our strategy is to employ the statistical constraint of
time dependent random multiplicative process to look for a solution of
the equations (\ref{basic}).  The theory of time dependent random
multiplicative process is in its infancy and only few exact results
have been obtained so far. One can wonder why we need a time dependent
stochastic process as a tool to describe equal time correlation
functions. The answer is that the shape of the correlation functions
is strongly dependent by the time dynamics \cite{bbs03}. The hope is
that, by using the dynamical scaling required by the Navier Stokes
equations, we can already obtain a good approximations to the real
solutions. Finally we want to comment on point (iii).  Shell models
provide the simplest model where to check our strategy and to compare
our physical ideas against clean numerical simulations in the
asymptotic regime of large Reynolds numbers.\\ Even with the
approximations defined in points (i)--(iii), the problem of computing
the universal anomalous scaling in turbulence is equivalent to solve a
functional equation, i.e. each equation (\ref{basic}) defines for any
order a constrain to be satisfied for the probability distribution. We
will limit ourselves to the lowest, non trivial, order such as to be able 
to calculation analytically as much as possible.\\ The paper
is organized in the following way.\\ In sect. (\ref{sect1.1}) we
briefly remind the basic properties of time-dependent random
multiplicative process. In section (\ref{sect2}) we address the
problem of anomalous scaling in Kraichnan shell models of passive
scalars. We present a re-derivation in the framework of stochastic
closure of an exact result for the anomalous scaling of fourth order
structure function.  Thus, we are able to connect the mathematical
notion of {\it zero-modes} with a {\it cascade mechanism} as given by
the time-dependent multiplicative process here used. In section
(\ref{sect3}) we extend the stochastic closure used fruitfully for the
passive scalar case to the fully non-linear model. We discuss at
length both similarities and differences between the two cases and we
present the first attempt to calculate the fifth-order scaling
exponents at varying the values of the free parameter in the shell
model.  Conclusions will follow in section (\ref{sect6}).
\section{Time-dependent random multiplicative process}
\label{sect1.1}
Let us first review the main ingredients of time-dependent
multiplicative process (TRMP) \cite{ben03}.  We introduce a set of
reference scales, $\ell_n = \ell_0 2^{-n}$, and a set of velocity
increments at scales, $\ell_n$: $\delta_n v \sim v(x+\ell_n)-v(x)$.
The basic idea of random {\it time-independent} multiplicative process
is to assume that statistical properties of $\delta_n v $ can be
obtained by 
\be \delta_{n+1}v = A_{n+1} \delta_n v,
\label{rmp_b}
\ee
where $A_i$ are i.i.d. (independent identically distributed) 
random variables with a time-independent --bare--
probability $P(A)$.   Time independent multiplicative
process as (\ref{rmp_b}) have been widely used in the past
to mimics the spatial distribution of velocity fluctuations
in turbulence and the multi-fractal energy dissipation measure 
\cite{fri95}. Recently, also a first attempt to match the stochastic
multiplicative model with the deterministic structure of the equation
of motion of a shell model of turbulence has been presented
\cite{ben93}. Despite the  success to reproduce the cascade
phenomenology, time-independent multiplicative process cannot capture
the subtle complexity inter-wined in the spatial and temporal behavior
of Navier-Stokes equations (in a Lagrangian reference frame).  For
example, multi-scale correlation functions of the kind, $\la \delta_n
v\delta_{n'}v \ra$ are well reproduced only asymptotically for large
scale separations, $n' \gg n$ \cite{ben98}.  The problem is that
simple time-independent random multiplicative processes do not take
in to account the time-dynamics, i.e. they are not constrained by the
equation of motion. To overcome this difficulties, recently, a new
class of time-dependent stochastic multiplicative process (TRMP) have
been proposed
\cite{bif99,ben03}. Basically, the idea is to mimics the temporal
constraints imposed by the structure of the Navier-Stokes eqs,
$\partial_t v \sim v \partial v$, by requiring that the
multiplicative structure (\ref{rmp_b}) is satisfied for the random time
interval $\tau_{n+1} = \ell_n/(\delta_n v)$. To build the temporal
dynamics we proceed as follows. 
We extract the instantaneous multiplier,  $A_n$, connecting 
the amplitudes of two velocity fluctuations
at adjacent scales, $\delta_n v=A_n \delta_{n-1}v$, 
with a given probability $P(A)$, independent form the scale, $\ell_n$,
 and keep it constant for a time interval 
$[t,t+\tau_n]$, with $\tau_n = \ell_n/(\delta_n v)$ being the local 
instantaneous eddy-turn-over time, 
 Thus, for each scale $\ell_n$, we introduce a time
dependent random process $A_n(t)$ which is piece-wise constant for a
 random time intervals $[t_n^{(k)},t_n^{(k)}+\tau_n]$, if $t_n^{(k)}$ 
is the time
of the $k$th jump at scale $n$. The corresponding 
 velocity field at scale $n$, in the time 
interval $t_n^{(k)} < t < t_n^{(k)}+\tau_n$,
is given by the simple multiplicative rule:
\begin{equation}
\label{time_dep}
\delta_n v(t)= A_n(t) \delta_{n-1}v(t_n^{(k)})
\end{equation}
What is important to notice is that at each jumping time,
$t_n^{(1)},t_n^{(2)},..,t_n^{(k)}..$ only the velocity field at the
corresponding shell, $n$ is updated, i.e. information across different
scales propagates with a finite speed.  In this way we reproduce the
phenomenology of the non-linear evolution of Navier-Stokes
dynamics: $\partial_t v \propto v \partial v$. Multipliers at different
scales, develop correlations through the time-dependency.  From now
on, we will denote with $\overline{\cdots}$ averages with respect to
the stochastic process and with $\la \cdots \ra$ averages over the
chaotic deterministic dynamics of the shell model. \\ A more detailed
numerical and theoretical analysis of the statistical properties of TRMP
can be found in \cite{ben03}.  The possibility to reproduce the {\it
single-time} statistical properties of TRMP by a Gibbs-like measure
has also been recently discussed in \cite{bbs03}.
%%%%%%%%%%%%%%%%%%%%%%%%%%%%%%%%%%%%%%%%%%%%%
\section{TRMP and the Kraichnan model}
\begin{center}
\begin{figure}
\epsfig{file= 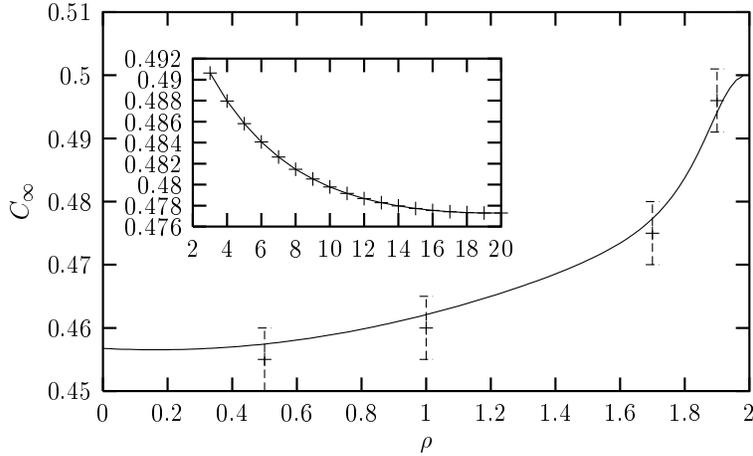,width=10.cm,keepaspectratio,clip}
\caption{Results for the asymptotic value $C_{\infty}$ of the fusion rules coefficients for
the Kraichnan model. For different values of  {$\xi$}, the result obtained by the analytic computation of the zero modes (continuous line) is compared with
the estimate ($+$) obtained  using the time dependent random multiplicative process (TRMP). Inset: plot of $\Gamma_{n}$ vs. $n$ in the case of the TRMP. The value of $\Gamma_{n}$ is multiplied by the factor $\frac{1}{2}$.}
\label{kraichnan}
\end{figure}
\end{center}

%%%%%%%%%%%%%%%%%%%%%%%%%%%%%%%%%%%%%%%%%%%%%%%%%
\label{sect2}
We start our work by understanding the relationship between the
Kraichnan shell model for passive scalar and time-dependent random
multiplicative process.  We first review the model and how the
computation of the anomalous exponents can be rigorously done in this
case.  The model is defined as follows \cite{wir96,ben97}.  Passive
increments are described on a discrete sub-set of wave-numbers
(shells), $k_n = k_0 \lambda^n$, by a complex variable $\theta_n(t)$,
which satisfy the equations ($n=1,2,\ldots,N$)
\begin{equation}
\left[\frac{d}{dt} + \kappa k_n^2\right] \theta_n (t) =  
  i [c_{n} \theta_{n+1}^* (t) u_{n}^*(t) +  b_n \theta_{n-1}^*(t)
  u_{n-1}^* (t)] + \delta_{1n} f(t)
\label{shellmodel}
\end{equation}
where the star denotes complex conjugation and $b_{n} = -k_{n}, c_{n}
=k_{n+1}$ are chosen such as to impose energy conservation in the zero
diffusivity limit. The inter-shell ratio must be taken $\lambda >1$.
Boundary conditions are defined as: $u_0=\theta_0=0$. The forcing term
$\delta_{1n} f(t)$ is Gaussian and delta correlated: $\la f(t)f(t')\ra
=F_1 \delta(t-t')$. It  acts only on the first shell. Kraichnan models
of passive advection   assume that each 
velocity variables, $u_n (t)$, is a   complex Gaussian and white-in-time
stochastic process, with a  variance which scales as: $\la
u_m (t) u_n^* (t') \ra =
\delta(t-t') \delta_{nm} d_m$, $ d_m= k_m^{-\xi}$.
The cross-correlation between the advecting velocity variables
and the passive variable can be rewritten in terms of passive
correlations only, when the velocity field is a white-in-time
Gaussian variable. Thus, all equations for all passive
structure functions are closed \cite{kra94,wir96,ben97}. 
The goal is to calculate the scaling exponents, $H(p)$ of
the p-th order passive structure functions as defined by:
$$
\la |\theta_n|^p \ra \sim k_n^{-H(p)}
$$
 We concentrate on the
non-perturbative analytic calculation of the fourth-order structure
function $P_{nn}= \langle (\theta_n \theta_n^*)^2 \rangle \propto
k_m^{-\zeta_4}$ (the lowest order with non-trivial anomalous scaling).
The closed equation satisfied by $P_{nq}=\langle (\theta_n
\theta_n^*)(\theta_q \theta_q^*)\rangle$, is, \begin{eqnarray}
\dot{P}_{nq} = (\delta_{1,n}  E_{n} +\delta_{1,q} E_{q})
 F_1 - \kappa
(k_n^2 +k_q^2) P_{nq} +\nonumber 
\quad \quad \quad \quad  \quad \quad\\ \left[-P_{nq} c_n^2d_n(
(1+\delta_{q,n+1}) + \lambda^{\xi-2} (1+\delta_{q,n-1}) )  +
 P_{n+1,q} c_n^2 d_n (1+\delta_{q,n})+ P_{n-1,q} b_n^2 d_{n-1}
(1+\delta_{q,n}) +( q \leftrightarrow n) \right];
\; \; \; \label{4} \end{eqnarray} where $E_n =
\langle\theta_n\theta_n^* \rangle = E_0 k_n^{\xi-2}$, i.e. the second
order scaling exponent is given by $H(2) = 2-\xi$. The above 
equation can be elegantly rewritten in the operatorial form:
\begin{equation}
\dot{P}_{nq} = {\cal I}_{nq,n'q'} P_{n'q'} + \kappa {\cal D}_{nq,n'q'} P_{n'q'}
+ {\cal F}_{nq},
\end{equation}
where we have explicitly separated the inertial, ${\cal I}$,
 from the dissipative, ${\cal D}$,
part of the linear operator and where the non-homogeneous term
composed by the forcing and by the second-order passive
structure functions  is summarized in the expression, ${\cal F}_{nq}$.
\\
It is useful to highlight in the two-scales correlation function,
$P_{n,n+l}$, the dependency from the scale separations by introducing
the set of variables, $C_l$:
\begin{equation}
P_{n,n+l} = C_l P_{n,n}.
\end{equation}
From basic scaling principle one may argue that 
the asymptotic scaling  behavior is given by 
the so-called fusion-rules \cite{eyi93,lvo96,bif99,ben98}:
\begin{equation}
P_{n,n+l} \sim  \frac{\la |\theta_{n+l}|^2 \ra}{\la |\theta_{n}|^2 \ra}
\la |\theta_{n}|^4 \ra \quad l \rightarrow \infty
\end{equation}
which means that $C_l \sim C_{\infty} k_l^{\xi-2}$ for $l$
positive and 
large enough. Similarly, for $l$ negative,  we may write:
\begin{equation}
P_{n,n-l} = D_l P_{n,n}
\end{equation}
where now the asymptotic behavior of $D_l$ feels the fourth-order
scaling behavior: $ D_l \sim D_{\infty} k_{l}^{-(\xi-2)-\rho_4}$ for
$l$ large enough, with $\rho_4 = H(4)-2H(2)$ being the anomaly of the
fourth order scaling exponent. The two sets of variables, $D_l$ and
$C_l$ are not independent. By introducing the notation, $x=\lambda^{\xi-2}$
and $R= \lambda^{\rho_4}$  one may rewrite both of
them as a function of a new set of variables, $\Gamma_l$
defined as: $C_l=\Gamma_l x^l$
and $D_l = \Gamma_l/ (xR)^l$
\cite{ben97}. The assumption that fusion rules are
satisfied is the only crucial point in computing the zero
modes. The existence of fusion rules implies that
correlation functions show scaling in the inertial range. \\
The infinite set of equations for the inertial-range {\it zero-mode} of 
(\ref{4}), ${\cal I}_{mq,m'q'} P_{m'q'}=0$,
 can  be rewritten in the following form:
\bea
\label{g0}
A_0(x,R) + B_{0,1}(x,R) \Gamma_1 = 0  &\quad q=n&\\
\label{g1}
A_1(x,R) + B_{1,1}(x,R) \Gamma_1 + B_{1,2}(x,R) \Gamma_2 = 0 &\quad n=q+1&\\
\label{gn}
B_{n,n-1}(x,R) \Gamma_{n-1} +B_{n,n}(x,R) \Gamma_{n}+ B_{n,n+1}(x,R)
\Gamma_{n+1} = 0 &\quad n > q+1&,
\eea
where the functions $A_0,A_1,B_{i,j}$ are known functions of $x$ and
$R=\lambda^{\rho_4}$. The computation of the zero modes means to find
out the numbers $R$ and $\Gamma_i$ which solves (\ref{g0}), (\ref{g1})
and (\ref{gn}).  Let us remark that for any given total 
number of shells, $N$, we
have $N+1$ equations and $N+2$ unknown which are given by the
$\Gamma_i$ for $i=1,\dots,N+1$ plus the parameter directly affected by
the fourth-order anomalous exponent, $R(\rho_4)$. Thus, it is
impossible to find a solutions unless some extra information is added
to our problem.  This information can be found by observing that for
large $n$ the functions $B$ appearing in (\ref{gn}) become constants
independent of both $x$ and $R$.  In the limit of large $n$, defining
$$ Z_n = \frac{\Gamma_{n+1}}{\Gamma_n} $$ one finds that equation
(\ref{gn}) can be rewritten as:
\be
Z_{n+1} = \Phi(Z_n)
\label{map}
\ee
where the explicit form of $\Phi$ is given in \cite{ben97}.  The map
(\ref{map}) possesses a fixed point $Z^*=1$ for large shell index, $n$,
which corresponds to the fact that $\Gamma_n$ reaches a plateau for
large $n$, i.e. to the fact that fusion rules are asymptotically
satisfied. The crucial point is to observe that $Z^*$ is a stable
fixed point for the inverse of $\Phi$, i.e. for
\be
Z_l = \Phi^{-1}(Z_{l+1}).
\label{mapinv}
\ee
The stability of $Z^*$ for (\ref{mapinv}) allows us to compute the
values of $Z_n$ for small $n$, i.e. we start with $Z_\infty = 1$ and
then we compute $Z_m$ by using (\ref{gn}) up to $m=2$. In this way we
can compute $Z_2$ as a function of $R$ and $x$. Thus equations
(\ref{g0}) and (\ref{g1}) become:
\bea
\label{g00} A_0(x,R) + B_{0,1}(x,R) \Gamma_1 = 0 \\
\label{g11} A_1(x,R) + B_{1,1}(x,R) \Gamma_1 + B_{1,2}(x,R) Z_2(x,R) \Gamma_1 = 0 .
\eea
Equations (\ref{g00}) and (\ref{g11}) have two unknowns, namely
$\Gamma_1$ and $R$, for two equations and, therefore, one can find a
solution.  The analytical 
solution turns out to be in perfect agreement with the numerics
both for the fourth order object here described 
\cite{ben97} and for higher order correlations \cite{an99}.
This ends the review of the analytical results previously obtained on the
model.\\ The solution of the Kraichnan shell model for passive scalar
provides us the rigorous computation of the zero modes.  We want now
to understand whether the computation of the zero modes can be pursued
by using the concept of time dependent random multiplicative
processes. In order to define a suitable TRMP to define the case of
the Kraichnan passive scalar we take the usual TRMP discussed in the
previous section for the updating of scalar fluctuations at two
adjacent scales:
\be
\theta_{n+1}(t) = A_{n+1}(t) \theta_n(t),
\label{rmp}
\ee
where $A_i$ are i.i.d.  random variables with a time-independent
probability $P(A)$.  The only difference with the TRMP for
the velocity field is that now the time upgrading of the multipliers
must satisfy the dynamical law: $\partial_t \theta \sim v \partial
\theta$. Thus, we need to update the multiplicative structure
(\ref{rmp}) at the random time interval $\tau_{n+1} = 1/(k_nu_n)$,
uncorrelated from the probability distribution of the multipliers
them-self (scalars are passive).  Moreover, because the advection field
is a Gaussian field with correlation functions proportional to
$k_n^{-\xi}$ one can deduce that $\tau_n$ is not a random time and
should be chosen as $\tau_n = k_n^{\xi-2}$.
One can show that such a class of TRMP predicts a non
trivial behavior of the fusion rules coefficients, $C_l,D_l$.
 Moreover, the
detailed behavior of the fusion rules coefficients are determined by
the spatial intermittency, i.e. the constrain $\partial_t \theta \sim v
\partial \theta$ between temporal and spatial scaling induced by the
Navier-Stokes structure in a Lagrangian reference frame is
satisfied. This is the crucial point we need to use to solve our
problem.  We can summarize our discussion in the following way. TRMP
provide us with a relationship between each fusion rule coefficients,
$C_l$, $D_l$ and the anomalous exponents. In this way we are building
a stochastic closure for eqs. (\ref{g0},\ref{g1},\ref{gn}).  Moreover,
the time and spatial dependencies of the stochastic process are
consistent with the structure of the deterministic equation of motion.
We want here to show that beside the exact method discussed before to
find the {\it zero-mode} also the stochastic closure trough the TRMP
works.\\ In the following we assume that the --bare-- probability
$P(A)$ is log normal. We are aware that log normal probability
distributions are not consistent with the anomalous scaling of
turbulent flows or shell models for large orders, even for the case of
the Kraichnan shell model.  However as far as we are interested to
compute $H(p)$ for rather small $p$, log normality is a reasonable
approximation which simplifies the analytical computations. Because we
know that $H(2)=2-\xi$, the probability distribution $P(A)$ depends
only on a single unknown parameter $\sigma$ which describes the
variance of the log normal fluctuations.  By using the exact solution
previously discussed, for each value of $\xi$ we can compute the value
of $H(4)$. Thus for each value of $\xi$ we can fix the parameters of
the log-normal distribution in order to reproduce the anomalous
exponent. We can next simulate the TRMP numerically and compute the
value of the fusion rules coefficients, $\Gamma_l$.  The most
sensitive test is made by comparing the prediction on the asymptotic
values of $\Gamma_l$ for large $l$ which we denote by $\Gamma_{\infty}$
(notice that $\Gamma_0=1$ by definition) as extracted from the
computation of the zero-mode and from the TRMP. \\
 Before doing a direct
comparison between the TRMP and the exact solution, we need to
discuss another  subtle point.  The definition of a random
multiplicative process shows an extra degree of freedom that is not
fixed neither by the scaling properties nor by the dynamical
scaling. To be more precise, in the case of the Kraichnan model, it is
possible to define $\T$ as: $$
\T_n(t) = g_n \prod_i^n A_i(t)
$$ where $g_n$ are i.i.d. random variable for any scale $n$.  Because
the probability distribution of $g_n$ does not depend on $n$, then the
scaling properties of $\T_n$ does not depend on $g_n$. However, the
fusion rules coefficients $\Gamma_l$ do depend on $g_n$. In particular
the quantity $\Gamma_{\infty}$ depends on $g_n$ as: $$
\Gamma_\infty (g=1) \rightarrow \Gamma_\infty (g) 
\frac{\la g^2\ra^2}{\la g^4\ra}
$$ Thus it seems that in our way to apply time dependent
multiplicative process to the Kraichnan shell model we are not able to
fix the fusion rules coefficients.  This is rather disappointing
because we start all our analysis by pointing out that the shape of
the fusion are determined by the time dynamics in a TRMP.  However,
the function $g$ and its probability distribution should not depend on
intermittency itself. In particular, it is relatively easy to compute
$g$ and its probability distribution for the Kraichnan model by
observing that for $\xi=2$ all scaling exponents $H(p)=0$ and $\rho_p
= 0$, as already observed in the work by Kraichnan \cite{kra94}.  Using
this informations in equations (\ref{g0},\ref{g1},\ref{gn}) 
we find that $\Gamma_1 =
1/2$.  Similarly, we can generalize this information for all fusion
rules coefficients.  This constrains can be satisfied only by a
suitable choice of $g_n$.  It turns out that in the Kraichnan shell
model this is equivalent to choose $g_n$ to be Gaussian. Thus the value
of $\Gamma_{\infty}$ should be multiplied by $2$ in order to compare it
with the TRMP. The comparison is done in figure
(\ref{kraichnan}).
As one can see the results are in extremely good agreement with the
exact solution. The above results provide us
with a complete and clear physical intuition of what a {\it zero-mode}
is. We have shown that the interpretation of anomalous scaling in terms
of {\it zero-modes} is fully compatible with the statistical properties 
of  multiplicative stochastic models. The only
missing brick was the importance of temporal dynamics. Anomalous
scaling as described by the zero-modes of the inertial operator for
the {\it simultaneous} $p$-th order structure functions is the outcome
of a physical {\it time-dependent} energy transfer from large scales to small scales.  Here we have shown
that a suitable closure based on TRMP is indeed sufficient to
calculate the zero-mode for fourth-order structure function, $\la
|\theta_m|^2 |\theta_q|^2 \ra$ in the inertial range.

%%%%%%%%%%%%%%%%%%%%%%%%%%%%%%%%%%%%%%%%%%%%%%
\section{Non-linear shell  model}
%%%%%%%%%%%%%%%%%%%%%%%%%%%%%%%%%%%%%%%%%%%%%%%
\label{sect3}
Here we want to understand if a result  similar to the one shown
in the previous section still holds true for the nonlinear shell models.
This  is important because: (i) we exploit the possibility to use TRMP in
the full non linear case; (ii) we can generalize the concept of zero
modes; (iii) we find out a way to compute the scaling exponents.  The
model we used is an improved version of the GOY model
\cite{yam87,jen91}, proposed in \cite{lvo98} (see also \cite{bif03}
for a recent review):
\be
\label{sabra}
\lp\frac{d}{dt}+\nu k_n^2\rp u_n = i\lp k_nu_{n+1}^*u_{n+2} + b k_{n-1}
u_{n+1}u_{n-1}^* + (1+b)k_{n-2}u_{n-2}u_{n-1}\rp +f_n,
\ee
where $u_n$ is a complex variable representing the velocity
fluctuations at wavenumber $k_n$, where $k_n=2^nk_0$.  Numerical
simulations show that the variables $u_n$ exhibit anomalous scaling
for $-1\le b \le 0$ , namely:
\be
S_p(n)\equiv \la\left|u_n^p\right|\ra \sim k_n^{-\zeta(p)},
\ee
where $\zeta(p)$ is a non linear function of $p$. Numerically it is
observed that the anomalous scaling behavior depends on the parameter
$b$ and it does not depend on the specific form chosen for the large
scale forcing $f_n$.
%By inspecting figure ??  we can safely state that TRMP is
%consistent, whithin the limitations and the approximations discussed
%so far, with the non linear dynamics of the Sabra model.
\subsection{Computing the scaling exponents}
\label{sect4}
By defining $Q_n(t) = u_n(t) u^*_n(t)$, we start by searching a solution of the
equation obeyed by the simultaneous fourth-order correlation in the limit of zero viscosity , i.e. neglecting dissipative effects:
\be
\frac{d}{dt} \la Q_mQ_n \ra = k_{n+1}\la Q_{m}W_{n+1}\ra +b k_{n}\la Q_{m}W_{n} \ra -(1+b)k_{n-1}\la Q_{m}W_{n-1}\ra +(n \leftrightarrow m)=0,
\label{smn}
\ee
where we have introduced  the flux-variable given by the third order
object $W_n = Im(u_{n+1}^*u_nu_{n-1})$. Equations (\ref{smn}) can be written as an infinite set
of linear equations for the fifth order correlation function: $\la
Q_nW_m\ra $. First, we can pick out the asymptotic behavior given by
the usual fusion-rule:
\be
\la Q_{n+l}W_n\ra  = D_l k_l^{-\zeta(2)}S_5(n) \;\;\; \la W_{n+l}Q_n\ra  = C_l k_l^{-\zeta(3)} S_5(n).
\label{frules}
\ee
Fusion rules are a general properties of the correlation functions in
turbulence, predicted by random multiplicative processes and verified
with very good accuracy in laboratory experiments \cite{lvo96,ben98}.
In particular, it is known that for large $l$, $C_l$ and $D_l$ are no
longer dependent on $l$.  

How to obtain informations on the behavior of $D_l$ and $C_l$? By
restricting ourselves to equal time correlation functions
(i.e. equations (\ref{basic})) there is no hope to close the problem
we are facing and it is impossible to get any useful information by
using equations (\ref{smn}).  In order to make progress, we now assume
as in section (\ref{sect2}), that the statistical properties of $u_n$
can be described in terms of a time dependent random multiplicative
process.  We will now employ the following approximations: (i) we use
$C_l$ and $D_l$ only for small $l$, i.e. $l \le 2$; (ii) for small $l$
we can assume that $C_l = D_l$.  Using these approximations we can
rewrite from (\ref{smn}) the equations regarding, $C_1$ and $C_2$ as
follows: \bea
\label{cc1} \la Q_nW_n\ra[C_1 + b -(1+b)C_1R^{-1}] = 0, \\ 
\label{ccn} \la Q_nW_n\ra[(1-(1+b)xR^{-1})C_{2} + b(1+x) C_1 +(Rx - (1+b))] = 0,
\eea
where we have introduced the short hand notation, $x=
\lambda^{-\zeta(2)}$, for the dependency on the second order exponents
and $R=\lambda^{\zeta(3)+\zeta(2)-\zeta(5)}$ for the dependency on the
fifth order anomaly, $\rho_5 =\zeta(3)+\zeta(2)-\zeta(5)$.  Also here,
as in the passive scalar, we have more unknowns than
equations. Precisely, once given the second order exponent, $\zeta(2)$,
we have two equations and three unknowns, the two fusion rules
coefficients for close-by shells, $C_1,C_2$ and the fifth-order
anomaly, $R(\rho_5)$.  Unfortunately one cannot follow the same
--winning-- strategy adopted for the passive scalar, because here the
equivalent of map (\ref{map}) is not stable for back iteration.  In
order to close the problem, we must provide informations on the value
$Z_2 = C_2 / C_1 $ as a function of $\rho_5$ and $\zeta(2)$. Here is where
we want to exploit the TRMP.

In order to apply our strategy, we first need to face the following
problem.  The structure of the equations we want to close (\ref{cc1})
deals explicitly with complex shell variables. Therefore one should
define two correlated random processes one for the amplitude, $|u_n|$,
and another for the phase of the velocity shell variable. Such a
strategy, although feasible, introduces new unknowns which need to be
fixed either by using the equations of motion or by using additional
informations.  To not increase the complexity of the problem we look
for a simpler and suitable approximation. The key point is that we
need to use TRMP just to obtain the quantity $Z_2 = C_2/C_1$, i.e. we
need to control the ratio
\be
{\cal R} = \frac{\la W_{n+2} Q_n \ra}{ \la W_{n+1} Q_n \ra }.
\label{ratio}
\ee
Defining $u_n = |u_n| exp(i\phi)$, we can write:
\be
{\cal R} = \frac{\la |u_{n+3}||u_{n+2}||u_{n+1}||u_n|^2 sin\Delta_{n+2}\ra}
{ \la |u_{n+2}||u_{n+1}||u_n|^3 sin\Delta_{n+1}\ra},
\label{ratio2}
\ee
where $\Delta_n = \phi_n+\phi_{n-1}-\phi_{n+1}$.  Expression
(\ref{ratio2}) tells us that, if the correlations between phases and
amplitudes is negligible, we can rewrite (\ref{ratio2}) 
as follows:
\be
{\cal R} = \frac{\la |u_{n+3}||u_{n+2}||u_{n+1}||u_n|^2 sin\Delta_{n+2}\ra}
{ \la |u_{n+2}||u_{n+1}||u_n|^3 sin\Delta_{n+1}\ra} \sim
\frac{\la |u_{n+1}|^3|u_n|^2 sin\Delta_{n+2}\ra}{\la |u_n|^5 sin\Delta_{n+1}\ra},
\label{ratio3}
\ee
where we have fused the shell variables at scales $n+3$ and $n+2$ with
shell variable at scale $n+1$ in the numerator and shell variables
$n+2$ and $n+1$ with shell at scale $n$ in the denominator.  The above
considerations can be formally stated by writing:
\be
\frac{\la W_{n+2}Q_n \ra}{\la W_{n+1} Q_n \ra}= {\cal K}(k_n,b)\frac{\la |u_{n+1}|^3|u_n|^2 \ra}{\la |u_n|^5\ra },
\label{defkappa}
\ee
where ${\cal K}(k_n,b)$ takes into account the correlation, if any,
between the phases ($\sin(\Delta_{n})$) and the amplitude of the shell
variables.  We expect that the quantity ${\cal K}(k_n,b)$, defined in
(\ref{defkappa}), does not depend on the scale (at least in the
inertial range) and might depend on the degree of intermittency,
i.e. on $b$.  In particular, if $ {\cal R}$ strongly depends on the
correlation between phases and amplitudes of the shell variables, then
${\cal K}$ should be strongly dependent on the free parameter, $b$,
entering in the definition of the non linear terms.

\begin{center}
\begin{figure}
\epsfig{file=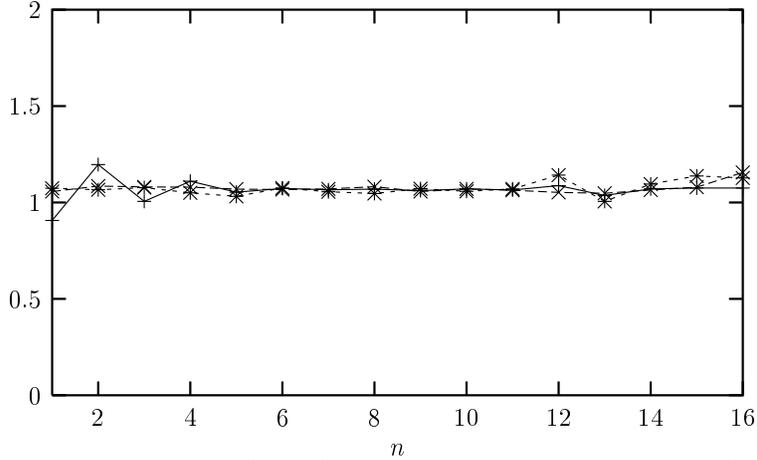,width=10.cm,keepaspectratio,clip}
\caption{Plot of ${\cal K}(k_{n},b)$
 vs $n$ computed from simulations of the shell model for the following values of the parameter $b$: $b=-0.4$ ($+$), $b=-0.6$ ($\times$), $b=-0.7$ ($\star$).
Only data in the inertial range are shown.}
\label{fig:kappa}
\end{figure}
\end{center}

The above discussion can be summarized by saying that the quantity
${\cal K}(k_n,b)$ is a direct measure of the importance of the
cross-correlations between shell amplitudes and phases with respect to
observable based only on shell amplitudes. In order to work out a
suitable strategy to apply TRMP as a statistical closure for the non
linear shell model, we only need that ${\cal K}$ is independent of
$b$.  Let us remark that such a requirement is not equivalent to a
"random phase approximation" (which would imply ${\cal K} = 1$). In
the following we shall assume that $\cal K$ is independent on
intermittency corrections, i.e. on $b$.
Our assumption is justified by the numerical results shown in figure
(\ref{fig:kappa}). As one can see, 
 the parameter ${\cal K}$ is indeed constant,
independent of both the shell index and of the intermittency intensity
as measured by the variation of the parameter $b$ in the equation of
motion.
Consequently, we may safely proceed with a simple TRMP based on
amplitudes only, using equation (\ref{defkappa}) with ${\cal K} =
const$ still to be determined.
Concerning the multipliers
distributions (\ref{rmp}), as in the case of the Kraichnan shell model, we
assume that $P(A)$ is log-normal.
Let us recall that the  exponents, $\zeta^{(s)}(p)$, measured from the scaling
of the stochastic signal, $\overline{|u_n|^p} \sim k_n^{\zeta^s(p)}$
do not coincide with the {\it bare} scaling exponents as estimated
by the instantaneous multiplicative process, $\zeta^{(b)}(p) = -log_{\lambda}
\la A^p \ra$, due to the correlation between the local
eddy turn over time and the velocity fluctuations, the time dynamics
{\it re-normalize} the spatial scaling \cite{ben03}. This is an extra
complication with respect to the passive case. Hereafter we always
refer to the bare exponents as $\zeta^{(b)}(p)$ and to those actually
measured on the stochastic signal as $\zeta^{(s)}(p)$.  We proceed by
performing the numerical estimate of the scaling properties of the
stochastic signal (\ref{time_dep}) at changing the parameters of the
{\it bare} log-normal distribution, $P(A)$, of the multipliers. Any
log-normal distribution is fixed by two -bare- parameters defining the
mean and the variance.  We fix the mean of the distribution such as
the third order exponent measured on the TRMP is consistent with the
$4/5$ law, $\zeta^{(s)}(3) =1$.  Now, we are left with only the
variance, $\sigma$, of the multipliers probability distribution,
$P(A)$, as a free parameter.  In order to have a control on the ratio:
$$ C_2/C_1 = {\cal K}\frac{\la |u_{n+1}|^3 |u_n|^2 \ra}{
\frac{\la|u_{n+1}|^3\ra}{\la|u_n|^3\ra} \la |u_n|^5 \ra} $$
we may estimate the unknowns multi-scale correlation functions appearing
in the r.h.s.  by using the TRMP at varying the log-normal
distribution:
\be
\frac{\la |u_{n+1}|^3 |u_n|^2 \ra}{
\frac{\la|u_{n+1}|^3\ra}{\la|u_n|^3\ra} \la |u_n|^5 \ra} \sim
\frac{\overline{|u_{n+1}|^3 |u_n|^2}^{(\sigma)}}{
\frac{\overline{|u_{n+1}|^3}^{(\sigma)}}{\overline{|u_n|^3}^{(\sigma)}}
\overline{|u_n|^5}^{(\sigma)}},
\label{fuscoeff}
\ee
where we have added a superscript $(\sigma)$ in the averages from the
TRMP to remind the dependency on the variance of the log-normal
distribution.\\
As a result we have a guess  on the ratio $C_2/C_1$ at varying $\sigma$
up to the still unknown constant ${\cal K}$.

\begin{center}
\begin{figure}
\epsfig{file=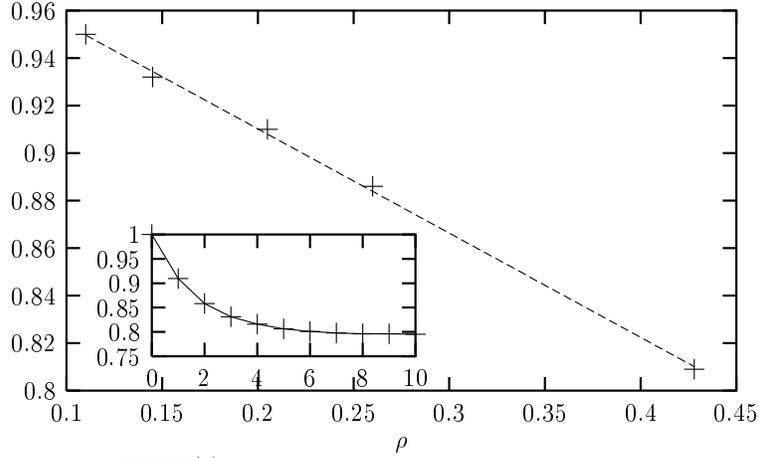,width=10.cm,keepaspectratio,clip}
\caption{Results for $\overline{|u_{n+1}|^3 |u_n|^2}^{(\sigma)}/
\frac{\overline{|u_{n+1}|^3}^{(\sigma)}}{\overline{|u_n|^3}^{(\sigma)}}
\overline{|u_n|^5}^{(\sigma)}$
 as a function of $\rho_5(\sigma) = \zeta^{(s)}(3) + \zeta^{(s)}(2) - \zeta^{(s)}(5)$.
 We plot the  data measured using the TRMP ($+$) and
the best linear fit $1-0.44\rho_5$. 
Inset: typical behavior of  the fusion rule coefficients, $C_{n}$, vs
 $n$ obtained from a TRMP stochastic signal.}
\label{fig:c1c2}
\end{figure}
\end{center}

The results of the numerical simulations are shown in figure
(\ref{fig:c1c2}) where we plot (\ref{fuscoeff}) as a function of $$
\rho_5(\sigma) = \zeta^{(s)}(3) + \zeta^{(s)}(2) - \zeta^{(s)}(5)
$$ in the TRMP.

\begin{center}
\begin{figure}
\epsfig{file=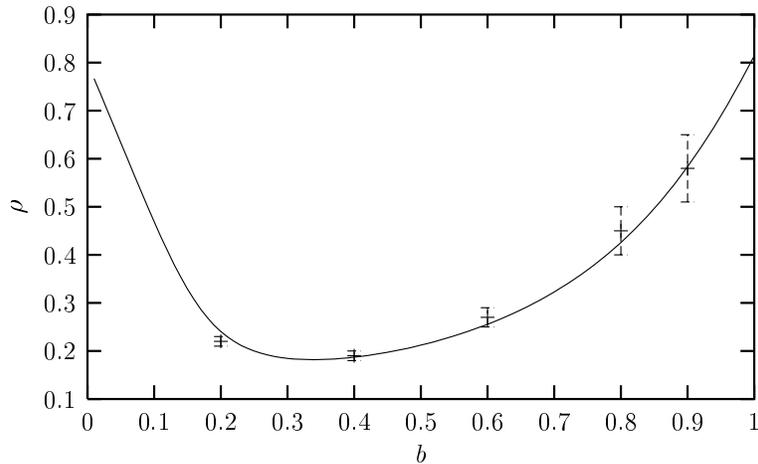,width=10.cm,keepaspectratio,clip}
\caption{The values of the fifth order anomaly  $\rho_5=\zeta(3)+\zeta(2)-\zeta(5)$ as a function of $b$ obtained from the TRMP closure approach (continuous line) are compared with the numerical estimate coming from simulations of the
shell model ($+$). } 
\label{fig:result}
\end{figure}
\end{center}

\begin{center}
\begin{figure}
\epsfig{file=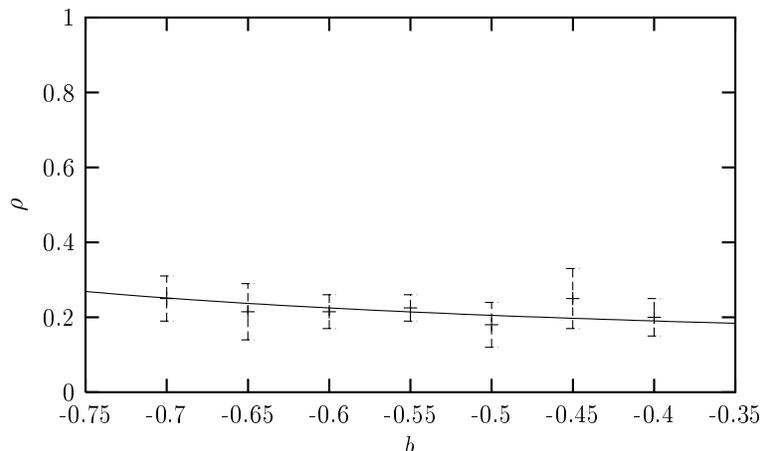,width=10.cm,keepaspectratio,clip}
\caption{The values for the fifth-order anomaly $\rho_5=\zeta(3)+\zeta(2)-\zeta(5) $ as a function of $b$ obtained from the TRMP closure approach along the curve $\lambda=\frac{1}{1+b} $  (continuous line) are compared with the numerical estimate coming from simulations of the
shell model ($+$).} 
\label{fig:res_lambda}
\end{figure}
\end{center}

As one can easily see from figure (\ref{fig:c1c2}), expression
(\ref{fuscoeff}) is extremely well 
fitted by a linear behavior:
\be 
\frac{\overline{|u_{n+1}|^3 |u_n|^2}^{(\sigma)}}{
\frac{\overline{|u_{n+1}|^3}^{(\sigma)}}{\overline{|u_n|^3}^{(\sigma)}}
\overline{|u_n|^5}^{(\sigma)}}
= 1 - 0.44 \rho_5(\sigma) \label{z2}.
\ee
This is the third equation, linking the unknowns in (\ref{cc1}) and
(\ref{ccn}) and closing the problem. It has been obtained by using the
TRMP.  This is not yet the end of the story. It is not enough to
plug the numerical result (\ref{z2}) into eqs. (\ref{cc1}) and
(\ref{ccn}) to close consistently the equations.  The problem is
connected to the possible presence of a {\it re-normalizing}
scale-independent stochastic variable in the multiplicative process
-- the $g_n$ variable already discussed for the passive scalar case.  We already 
discussed that the presence of such a scale-invariant distribution
changes only one  overall constant in the multi-scale
behavior. Moreover, we already know that another unknown constant
overall, ${\cal K}$, shows up due to the phases' statistics.  Summing
the two effects, we can assume that the {\it true} ratio $ C_2/C_1$
which must be plugged in the equation of motion can be estimated by
the result of the TRMP (\ref{z2}) plus a multiplicative, unknown,
constant, ${\cal D}$ independent on the intermittency of the model:
\be
\frac{C_2}{C_1} = \lp 1 - 0.44 \rho_5(\sigma) \rp {\cal D}.
\label{z2g}
\ee
\\
 Using equations (\ref{cc1},\ref{ccn},\ref{z2g}), we can
compute the fifth-order anomaly, $\rho_5= log_{\lambda}(R)$,
 by solving the system of three equations
in three unknown, $C_2,C_1,\rho_5$:
\bea
\label{f1} C_1 + b -(1+b)C_1R^{-1}(\rho_5) = 0 \\ 
\label{f2} (1-(1+b)xR^{-1}(\rho_5))C_2 + b(1+x) C_1 +(R(\rho_5)x- (1+b)) = 0 \\
\label{f3} C_2/C_1 = {\cal D} (1-0.44\rho_5)
\eea
To our knowledge, there are no simple theoretical arguments which can
be used in order to fix the value of ${\cal D}$. We fix it by assuming
that for $b=-0.4$ we recover the value of $\rho_5$ computed in the
numerical simulations.  It turns out that ${\cal D} = 0.85$.  We can
then compute $\rho_5$ for all values of $b$ in the range $-1<b<0$.  In
Figure (\ref{fig:result}) we show the computation of $\rho_5$ obtained
by using (\ref{f1}-\ref{f3}) together with the numerical estimate of
$\rho_5$ obtained by simulations of the shell model. As one can see the
results are in very good agreement with the numerical data {\em for the
whole range of b}.
In order to validate our results, we have compared the estimate of the
anomalous anomaly $\rho_5$ for the values of $b$ and $\lambda$
corresponding to the curve
\be
\lambda = \frac{1}{1+b}.
\label{hel}
\ee
 The
curve (\ref{hel}) is determined by the requirement that the second in-viscid
invariant, beside the total energy, $\sum_n {u_n}^2$, keeps the
physical dimensions of Helicity, $\sum_n (-)^n k_n |u_n|^2$
\cite{bif95b,ben96}.  It is known \cite{kad95} that along this curve
intermittency stay constant. 
Numerical estimate of (\ref{z2}) for different values of 
$\lambda$ do not show any appreciable difference with respect to 
what plotted in figure (\ref{fig:c1c2}).
 Thus, we can still use (\ref{f3}) as a numerical
estimate of $C_2/C_1$ with {\em the same value of ${\cal D}$}.
In fig.~(\ref{fig:res_lambda}) we show the comparison
between the results of the closure on the special curve (\ref{hel}) and the 
values estimated by numerical simulations.
Again, the stochastic closure
works perfectly, allowing for a precise determination of
the anomaly along this curve. 

\section{Conclusions and discussions}
\label{sect6}
In this paper we have discussed a possible strategy to compute the
scaling exponent of a non linear shell model.  The main idea of this
strategy is to assume that the statistical properties of the shell
variables $u_n$ 
 can be described in terms of a time dependent random multiplicative
process.  A mathematical way to summarize this idea is the following.
Let us define the random variable $A_n$ as:
\be
|u_{n+1}| = A_n |u_n|.
\ee
Let us also assume that $A_n = \lambda^{h_n}$, 
where the random variables $h_n$ 
fluctuates with probability
distribution $P_n = Z_n \cdot (\exp{-V(h_n)})$. 
If we neglect time dynamics 
the probability distribution of the shell variables is given by
the product $\Pi_k P_k$. Time dependent random 
multiplicative process provides well defined 
correlations among the random variables for
different scales. One can therefore write:
\be
P(u_1,u_2,...u_n,...) = 
Z\cdot \exp{\left[-\Sigma_n V(h_n) + \Sigma_{i,j} R_{ij}h_ih_j \right]}.
\ee
Once ${V}(h_n)$ is defined, the time dependency selects an unique
value of $R_{i,j}$. Formally we can write $R_{i,j} = \Gamma_{i,j}[V]$.
It follows that in order to compute the scaling exponents we need to
compute ${V}(h_n)$. The role of $R_{i,j}$ is crucial because it
determines the full shape of the coefficients needed to compute the
fusion rules. Using the time average equation of motions for the
structure functions, we are therefore able to obtain a functional
equation for $V$, whose solutions provides the anomalous scaling
exponents of the shell model. In order to understand whether our
strategy is providing reasonable results, we have assumed that ${
V}(x)$ is a quadratic function of $x$. Thus, because $\zeta(3)=1$, we
have only one unknown to be computed corresponding to the quadratic
non linearity. With this assumption, the functional equation for ${V}$
reduces to an equation for one unknown (the quadratic non linearity)
which we have solved.  The results we obtain are in very good
agreement with the numerical simulation of the shell model.\\ We want
to stress that the same procedure can be applied for the passive
scalar advected by a random velocity field (the Kraichnan model) with
extremely good results and without any ad hoc approximation.\\ We
want to highlight few points which we believe are true independent of
the approximations we did in this paper.  (i) The computations of the
scaling exponents are feasible if the fusion rules coefficient are
known as a function of intermittency (i.e. the function $D(h)$ in the
multi-fractal language).  (ii) The fusion rules coefficient depends on
intermittency because of time dynamics.  (iii) Time dependent
random multiplicative processes are consistent with the dynamical
deterministic structure of the equation of motion and 
provide an useful tool to compute fusion
rules coefficients.\\ Nevertheless, it is not clear yet, 
if the stochastic process
successfully applied here to close the
equation of motion of fourth-order velocity correlation is
also the optimal solution for higher order correlation functions.
In other words, one has to face also the possibility that different
fluctuations (controlling higher order correlation functions) are described
by different stochastic processes. 

In applying our strategy to the shell model we have performed a number
of approximations. In particular we consider an important point to
generalize our approach in order to properly take into account the
phase dynamics in the shell model. Also, the approximation $C_l = D_l$
should be considered as a first order of approximation in order to
develop a systematic theory.  We also want to highlight two important
topics for future research. First of all, we think that it is
important to apply our method in the case of the Kraichnan model in
two or more space dimensions. In order to perform such a task we need
to develop the field theory of time dependent random multiplicative
processes.  Also, we need to have a more detailed analytical control
of time dependent random multiplicative processes, following the ideas
already discussed in \cite{ben03,bbs03}.

This research was supported by the EU under the Grants
No. HPRN-CT 2000-00162 ``Non Ideal Turbulence''.


\begin{thebibliography}{10}

\bibitem{fri95}
U. Frisch, {\em Turbulence: The legacy of A.N. Kolmogorov} (Cambridge
  University Press, Cambridge, 1995).

\bibitem{ben93b}
R. Benzi, S. Ciliberto, R. Tripiccione, C. Baudet, F. Massaioli, and S. Succi,
  Phys. Rev. E {\bf 48},  R29  (1993).

\bibitem{fgv01}
G. Falkovich, K. Gaw\c{e}dzki, and M. Vergassola, Rev. Mod. Phys. {\bf 73},
  913  (2001).

\bibitem{bif03}
L. Biferale, Annu. Rev. Fluid. Mech. {\bf 35},  441  (2003).

\bibitem{ben84}
R. Benzi, G. Paladin, G. Parisi, and A. Vulpiani, J. Phys. A {\bf 17},  3521
  (1984).

\bibitem{ben93}
R. Benzi, L. Biferale, and G. Parisi, Physica D {\bf 65},  163  (1993).

\bibitem{bif99}
L. Biferale, G. Boffetta, A. Celani, and F. Toschi, Physica D {\bf 127},  187
  (1996).

\bibitem{ben03}
R. Benzi, L. Biferale, and F. Toschi, J. Stat. Phys.  (2003), in press,
  nlin.CD/0211005.

\bibitem{bbs03}
R. Benzi, L. Biferale, and M. Sbragaglia, J. Stat. Phys.  (2003), submitted,
  nlin.CD/0211005.

\bibitem{ben98}
R. Benzi, L. Biferale, and F. Toschi, Phys. Rev. Lett. {\bf 80},  3244  (1998).

\bibitem{wir96}
A. Wirth and L. Biferale, Phys. Rev. E {\bf 54},  4982  (1996).

\bibitem{ben97}
R. Benzi, L. Biferale, and A. Wirth, Phys. Rev. Lett. {\bf 78},  4926  (1997).
\bibitem{an99} 
 K.H. Andersen an P. Muratore-Ginanneschi,  Phys. Rev. E  {\bf 60} 
6663 (1999).
\bibitem{kra94}
R. Kraichnan, Phys. Rev. Lett. {\bf 72},  1016  (1994).

\bibitem{eyi93}
G. Eyink, Phys. Lett. A {\bf 172},  355  (1993).

\bibitem{lvo96}
V. L'vov and I. Procaccia, Phys. Rev. Lett {\bf 76},  2898  (1996).

\bibitem{yam87}
M. Yamada and K. Ohkitani, J. Phys. Soc. Jpn. {\bf 56},  4210  (1987).

\bibitem{jen91}
M.~H. Jensen, G. Paladin, and A. Vulpiani, Phys. Rev. A {\bf 43},  798  (1991).

\bibitem{lvo98}
V.~L. V, E. Podivilov, A. Pomyalov, I. Procaccia, and D. Vandembroucq, Phys.
  Rev. E {\bf 58},  1811  (1998).

\bibitem{bif95b}
L. Biferale and R. Kerr, Phys. Rev. E {\bf 52},  6113  (1995).

\bibitem{ben96}
R. Benzi, L. Biferale, R. Kerr, and E. Trovatore, Phys. Rev. E {\bf 53},  3541
  (1996).

\bibitem{kad95}
L. Kadanoff, D. Lohse, J. Wang, and R. Benzi, Phys. Fluids {\bf 7},  617
  (1995).

\end{thebibliography}
\end{document}